\documentclass[%
 reprint,
 amsmath,amssymb,
 aps,
]{revtex4-2}

\usepackage{graphicx}
\usepackage{dcolumn}
\usepackage{bm}
\usepackage{lineno}
\usepackage{natbib}
\begin{document}

\preprint{APS/123-QED}

\title{Testing hadronic and photo-hadronic interactions as responsible for UHECR \\ and neutrino fluxes from Starburst Galaxies}

\author{Antonio Condorelli}
 \altaffiliation[ ]{condorelli@ijclab.in2p3.fr}
\affiliation{%
 Laboratoire de Physique des 2 Infinis Irène Joliot-Curie, CNRS/IN2P3, Université Paris-Saclay, France \\
 }
 \affiliation{
 Gran Sasso Science Institute,
Via F. Crispi 7, 67100, L’Aquila, Italy\\
}%
\author{Denise Boncioli}
\affiliation{
Dipartimento di Scienze Fisiche e Chimiche, Universit\`a degli Studi dell'Aquila, via Vetoio, 67100, L'Aquila, Italy\\}
\affiliation{
INFN/Laboratori Nazionali del Gran Sasso, via G. Acitelli 22, 67100, Assergi (AQ), Italy\\
}%
\author{Enrico Peretti}
\affiliation{%
Niels Bohr International Academy, Niels Bohr Institute, University of Copenhagen, Blegdamsvej 17, DK-2100 Copenhagen, Denmark \\
}
\author{Sergio Petrera}
\affiliation{%
Gran Sasso Science Institute,
Via F. Crispi 7, 67100, L’Aquila, Italy \\ }
\affiliation{
INFN/Laboratori Nazionali del Gran Sasso, via G. Acitelli 22, 67100, Assergi (AQ), Italy\\
}

\date{\today}

\begin{abstract}
We test the hypothesis of starburst galaxies as sources of ultra-high energy cosmic rays and high-energy neutrinos. The computation of interactions of ultra-high energy cosmic rays in the starburst environment as well as in the propagation to the Earth is made using a modified version of the Monte Carlo code {\it SimProp}, where hadronic processes in the environment of sources are implemented for the first time. Taking into account a star-formation-rate distribution of sources, the  fluxes of ultra-high energy cosmic rays and high-energy neutrinos are computed and compared with observations, and the explored parameter space for the source characteristics is discussed. We find that, depending on the density of
the gas in the source environment, spallation reactions could exceed the
outcome in neutrinos from photo-hadronic interactions in the source
environment and in the extra-galactic space.
\end{abstract}
\keywords{Suggested keywords}
\maketitle


\section{Introduction}

One of the most exciting astrophysical discoveries of the last century is the existence of a diffuse flux of cosmic particles  extending in energy up to { $\sim 10^{20} \, \rm eV$,} { an energy range} greatly exceeding every Earth-based accelerator. 
Decades of observations have allowed us to explore its spectral behavior and composition in terms of atomic nuclei \citep{Coleman:2022abf}. 
However its nature and origin remain a mystery, thereby making the puzzle of ultra-high energy cosmic rays (UHECRs) one of the most intriguing open questions of modern astrophysics. 
In order to provide an answer to such a question, the Pierre Auger Collaboration \citep{PierreAuger:2015eyc} has published a study \citep{Aab:2018chp}, recently updated in \cite{PierreAuger:2022axr}, in which the correlation between UHECRs at the highest energies and source catalogues is explored. 
 In particular,  a strong correlation  has been found (4.2$\sigma$, foreseen to become 5$\sigma$ in 2026)  between the arrival directions of UHECRs and the coordinates of the starburst galaxies (SBGs) in the catalogue, even if the contribution from SBGs is summed to an isotropic background; these results supported the idea of SBGs as potential class of sources for UHECRs. 

If UHECRs  were produced in the  most active regions of SBGs, known as starburst nuclei (SBNi), one of the  key points to investigate would be the impact of the  starburst environment on the UHECR interactions. 
Indeed, in several works, both in the case of generic parameters describing the sources \citep[see e.g.][]{Allard_Proterhoe,Unger:2015laa,Supanitsky,Muzio:2019leu,Muzio:2021zud} and for specific source classes \citep[see e.g.][]{Globus:2015xga,Biehl:2017zlw,Biehl:2017hnb,Boncioli:2018lrv} it was shown how the post-processing of UHECRs via photo-disintegration of CR nuclei in the environment surrounding an hypothetical source can qualitatively explain the UHECR spectrum and composition across the ankle, meaning the flattening of the spectrum near $5\times 10^{18}$ eV \cite{PierreAuger:2020qqz}. 

In these models, hereafter referred to as ``source-propagation models", the photo-disintegration process acts as a high-pass filter allowing the highest energy cosmic-ray nuclei to escape unscattered whereas the lowest energy ones are disintegrated inside the source region, thereby generating a pile-up of nucleons with energy scaling as $1/A$, being $A$ the mass of the nucleus injected in the acceleration region. 
Particles escaping the source environment are then propagated through the intergalactic medium and  finally the obtained diffuse fluxes are compared to the experimental data at Earth. 
Recently, hadronic interactions with the source environment have been also considered for a generic source, showing that they can contribute to the same escape effect even though with less efficiency \citep{Muzio:2021zud}.

Diffuse fluxes of gamma rays (up to TeV energy) and HE neutrinos (up to PeV energy) have been observed respectively by Fermi-LAT \citep[see][]{Fermi-LAT<820} and IceCube \citep[see][and references therein]{IceCube2020_LAST}.
Starburst galaxies have been already proposed as potential candidates for such fluxes \citep[see e.g.][]{Tamborra-Ando-Murase:2014,Bechtol-Ahlers:2015,Sudoh,Peretti-2,Ambrosone2020,Roth-nature,Owen_diff1}, however a detailed modeling of UHECR interactions in their environment has not been deeply explored yet, mostly due to the current lack of an acceleration model able to inject particles at the highest energies.
In fact, while cosmic accelerators such as powerful supernova remnants \citep[see e.g.][and references therein]{Cristofari:2020mdf,Cristofari_PeVatrons} or young massive star clusters \citep[see e.g.][]{Morlino_2021} can possibly accelerate particles up to PeV, it is not clear whether and where the acceleration in the EeV range can take place in a starburst environment or in the associated wind bubbles \citep[see e.g.][]{Romero_wind,Muller-Romero,Anchordoqui_SBG,Peretti-wind}.
The current lack of a detailed theory for particle acceleration makes a phenomenological investigation of UHECRs in starburst environment timely and key for understanding the main properties of such particle population.

In this work, we assume that SBNi are capable to power UHECRs and, for the first time, we use a source-propagation model
to derive the UHECR and high-energy neutrino fluxes from these sources.
We develop an extension of the Monte Carlo code \textit{SimProp} \citep[see][]{Aloisio:2012wj,Aloisio:2015sga,Aloisio:2016tqp} to simulate the behavior of UHECRs and study the multimessenger implications in terms of associated High Energy (HE) neutrino flux, focusing on the role of the hadronic and photo-hadronic interactions in these environments.

The paper is organized as follows: in Section \ref{Sec: 2-Starburst} we introduce SBGs as potential sources of UHECRs, highlighting their common properties and detailing the way we compute the interactions of UHECRs in this environment; in Section \ref{Sec: 3-Experimental_data} the parameter space is discussed and the method to search for the best configuration of parameters is described; in Section \ref{Sec: 5-Results} the comparison to UHECR data is presented, together with a discussion on the effects of varying some parameters at the source on the observables at Earth. The expected neutrino fluxes associated to the chosen set of parameters at the source are also presented, separating the contributions coming from different interactions in the sources, as well as the neutrinos expected to be produced in the extra-galactic propagation. We finally draw our conclusions in Section \ref{Sec: 6-Discussion-Conclusions}.


\section{UHECR interactions in Starburst Nuclei}
\label{Sec: 2-Starburst}

Starburst galaxies  are unique astrophysical objects characterized by an intense star forming activity which can be as high as $\sim 10-10^3 \, \rm M_{\odot} yr^{-1}$ \citep[see][]{Gao-SB-SF}. 
As detailed in \cite{Rieke}, the higher the star formation rate (SFR) the greater the infrared luminosity, and according to \cite{Mannucci-SB-IR}, a corresponding increment of the rate of supernovae (SNe) ($\mathcal{R}_{\rm SN} \sim 0.1 \div 1 \, \rm yr^{-1}$) is often observed. 
{ Such an enhanced rate of SNe makes SBGs very efficient cosmic ray (CR) factories and, in turn, this connection between SNe and SFR results in correlations observed between the non-thermal luminosity and the SFR} \citep[see e.g.][and references therein]{Kornecki_2020,Kornecki_2022}. 

In several SBGs,  most of the star formation is observed to be localized in SBNi located in the central part of the galaxy with typical extension ranging from a few hundred parsecs to kiloparsecs. 
Due to the intense activity in such a compact environment,
their interstellar medium (ISM) is naturally expected to be highly perturbed with a strong level of turbulence \citep[see][]{Peretti:2018tmo}. 

The SBNi environment exhibits extreme conditions such as a  gas density as high as (or higher than) $n_{\rm ISM} \sim 10^2 \, \rm cm^{-3}$ \citep[see also][]{Kennicutt:1997Relation,Forster-Schreiber_n}, magnetic field ($B$) at the level of $\sim 0.1 - 1 \, \rm mG$  \citep[see][]{Thompson_B} and infrared photon density ($U_{\rm RAD}$) often higher than $10^3 \, \rm eV \, cm^{-3}$. 
{ In addition,} the superposition of several SNe and the intense star forming activity could favor the conditions to launch a powerful wind with estimated velocity ($v_{\rm W}$) of about $\sim 10^2 - 10^3 \, \rm km \, s^{-1}$ \citep[see e.g.][]{CC85,Seaquist_SB_wind,Engelbracht_wind,Strickland_wind}.

The high level of turbulence { expected for the SBN} environment suggests that PeV and sub-PeV CR protons might lose a consistent part of their energy {through} proton-proton (pp) interactions before  {being able to escape,} {mostly due to the advection in the wind} \citep{Yoast-Hull_M82_2013,Wang_2018,Peretti:2018tmo}. 
Recent investigations proposed that CRs can be additionally accelerated up to $\sim 10^2 \, \rm PeV$ at shocks in the wind bubbles inflated by the starburst activity in the SBN \citep{Romero_wind,Peretti-wind}, whereas in \cite{Anchordoqui_SBG} it is argued that in the same conditions energies up to $\sim 10^2 \, \rm EeV$ could be achieved. 
These HE particles would still lose part of their energy via $pp$ and proton-gamma ($p\gamma$) interactions on the diluted photon field of the SBN, but their energy will efficiently allow them to diffuse away from the starburst surroundings.

{ Beside the acceleration sites directly connected to the starburst activity, such as supernova remnants, long gamma ray bursts \citep[][]{Ghisellini_2008}, massive star winds \citep[see e.g.][]{Morlino_2021} and the starburst-driven wind \citep[][]{Peretti-wind}, there can be additional phenomena responsible for the injection of HE particles in the SBN environment.    
In particular, star forming activity is often observed to be coexisting with an activity of the supermassive black hole (SMBH) hosted in the galaxy core. 
In fact, active galactic nuclei (AGN) can launch relativistic jets} \citep[see][and references therein]{Alves_Batista_2019} { and spherically symmetric sub-relativistic winds}\citep[see e.g.][]{Lamastra1,Wang-Loeb2017,Liu2018,Lamastra2} { where HE particles can be accelerated or possibly reaccelerated} \citep[][]{Caprioli_Espresso}. 

In this work, assuming that CR nuclei accelerated in the SBN environment up to the highest energies, we focus our attention on the multi-messenger implications of such particle population in terms of UHECRs and HE neutrinos.

\begin{figure}

	\includegraphics[width=\columnwidth]{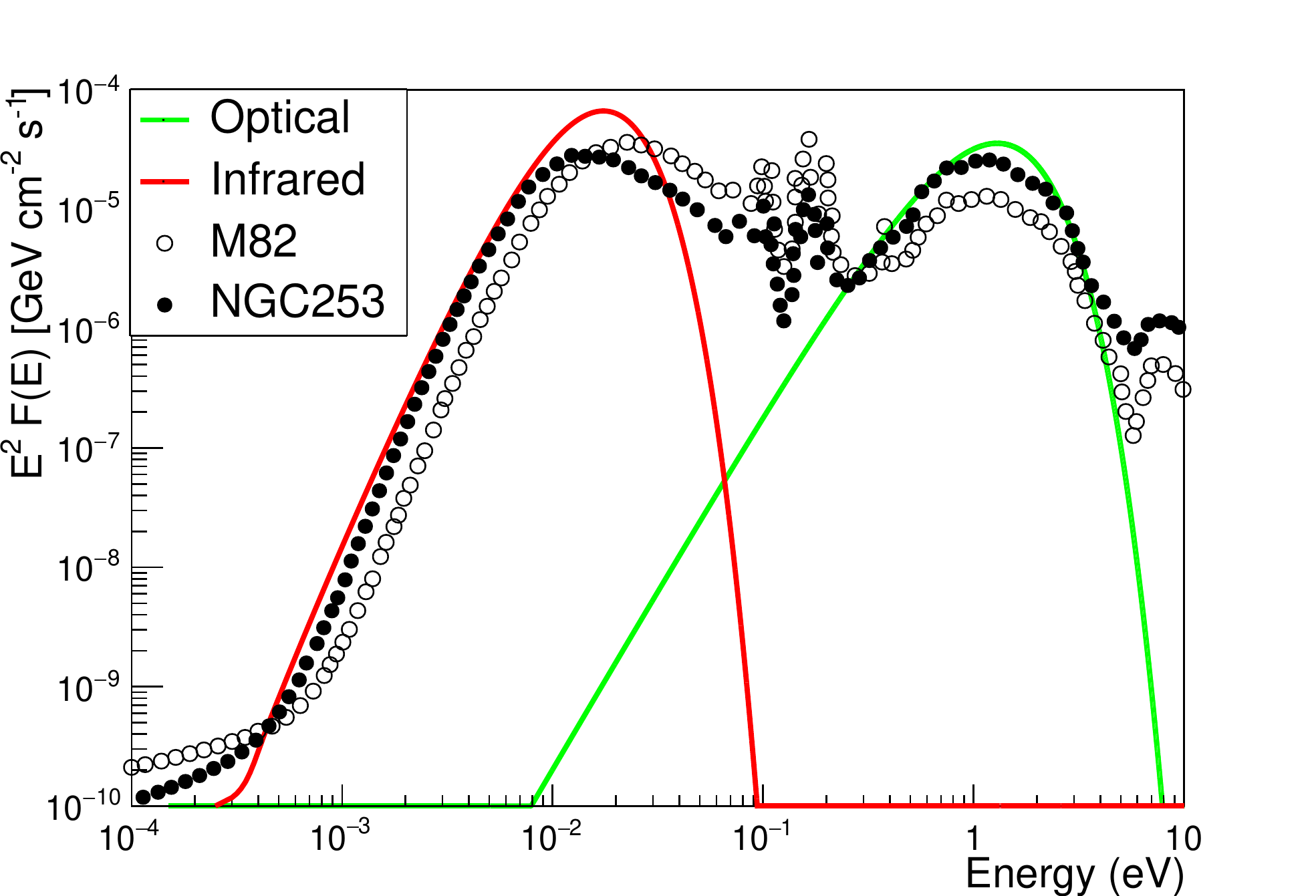}
    \caption{Photon spectrum of the prototype SBG, as inspired by \citep{Peretti:2018tmo}: thermal dust modified black bodies (red line) and optical star black body (green line). The black points refer to the measurements from \citep{Galliano2007StellarEE} for two different SBG: M82 and NGC253. }

    \label{fig:M82_prototype}
\end{figure}
\begin{table}

    \centering
\begin{tabular}{p{3.0cm}|p{1.3cm}} 
{Parameter}  & {Value} \\ \hline
{$R \ \rm (pc)$}   & 225   \\
{$B \ \rm (\mu G)$}   & 200   \\
{$v_{\rm wind}$} \ ($\rm km \ s^{-1}$)   & 500   \\
{$n_{\rm ISM} \ (\rm cm^{-3})$}   & 125   \\
${U_{\rm eV \ cm^{-3}}^{\rm FIR} \ \bigg[\dfrac{kT}{\rm meV}\bigg] }$   &    1958 [3.5]  \\
${U^{\rm OPT}_{\rm eV \ cm^{-3}} \ \bigg[\dfrac{kT}{\rm meV}\bigg] }$   &    2936 [332.5]  \\

\hline
\end{tabular}
 \caption[Parameter space for prototype case]{Parameters 
 of the prototype SBG.}
    \label{tab:Reference_SBG}

\end{table}

\subsection{Interactions and escape from starburst environment}

{The low energy photon field of SBGs is complex and characterized by a superposition of several thermal components of different nature ranging from the far infrared (FIR) up to the optical (OPT) and ultraviolet (UV) \citep[see e.g.][for additional details]{Galliano2007StellarEE}. 
In particular, the most prominent spectral components are 1) a blackbody associated to the starlight peaking at $\varepsilon_{\rm opt} \simeq 1 \, \rm eV$ and 2) a second thermal component peaking at $\varepsilon_{\rm IR} \simeq 10 \, \rm meV$ resulting from the reprocessing of the UV light by interstellar dust.}
{ We assume a stereotypical low-energy spectral energy distribution (SED) approximated by these two thermal components.}
In order to study the environment surrounding the SBN and how it impacts on UHECRs, in our work it was chosen to adopt a prototype SBG, i.e., a SBG with parameters listed in Tab.~\ref{tab:Reference_SBG}. The photon spectrum for our prototype SBG is shown in Figure \ref{fig:M82_prototype}  where it is compared with the spectra of two nearby starbursts: M82 and NGC253 \citep[][]{Galliano2007StellarEE}.\\
In the following, typical timescales for photo-hadronic and hadronic interactions of CR particles in the SBN are described, as computed from a modified version of the Monte Carlo code \textit{SimProp}.\\

%

\begin{figure}
	\includegraphics[width=\columnwidth]{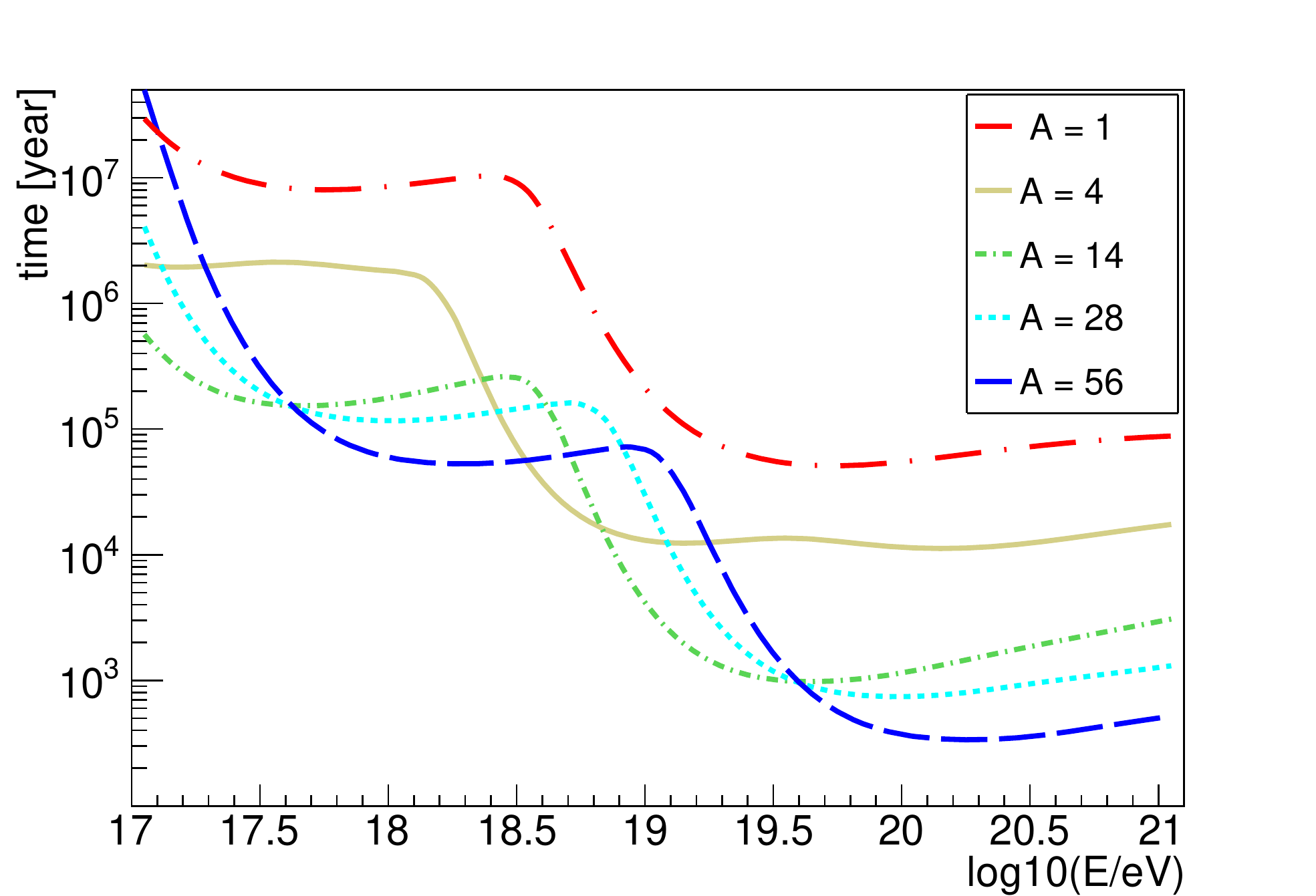}
    \caption{Timescales for photo-interactions in our prototype SBG for five different injected CR nuclear species as indicated in the legenda (see Tab. \ref{tab:Reference_SBG} for details).}
    \label{fig:M82_photo_time}
\end{figure}

Under the assumption of a monochromatic photon field of photon density $n_{\gamma}$, the typical interaction rate  between a relativistic atomic nucleus ($A$) and a low energy photon is approximately $\tau_{A\gamma}^{-1} \simeq c \sigma_{A \gamma} n_{\gamma}$, where $\sigma_{A \gamma}$ represents the cross section of the process.
If  a more realistic photon density is considered and the dependence of the cross section on the energy is taken into account, the interaction rate reads:
\begin{equation}
\dfrac{dN_\text{int}}{dt} = \dfrac{c}{2\Gamma}\int_{\epsilon'_\text{th}}^\infty \sigma_{A\gamma}(\epsilon')\epsilon' \int_{\epsilon'/2\Gamma}^\infty \dfrac{n_\gamma (\epsilon)}{\epsilon^2} \,d\epsilon\,d\epsilon'
\label{eq: interaction_time}
\end{equation}
where 
$\Gamma$ is the Lorentz factor of the interacting nucleus. Note that primed symbols (e.g. $\epsilon'$) refer to quantities in the nucleus rest frame, whereas unmarked symbols refer to quantities in the laboratory frame.
The interaction timescales, corresponding to the inverse of Eq.~\eqref{eq: interaction_time}, are shown in Fig.~\ref{fig:M82_photo_time} for different nuclear species, for which the following reactions are taken into account: photo-production of electron-positron pairs, nuclear photo-disintegration and photo-pion production. The dip at low energies corresponds to the photo-disintegration on the OPT component, while at higher energies the interaction timescale is dominated by the FIR peak. The effect of the other reactions mentioned above is smaller with respect to the photo-disintegration.\\ 

Though spallation processes between the CR nuclei and gas environment have negligible impact in the extra-galactic medium, their role is remarkable in the ISM of SBNi given the typical densities associated to
active star-forming regions. The timescale for the spallation process reads:
\begin{equation}
\tau_{\rm spal} = \frac{1}{n_{\rm ISM} \, \sigma \, c },
\label{eq:time_spal}
\end{equation}
where $n_{\rm ISM}$ is the  ISM gas density in the  SBN environment. 
This process has been implemented in \textit{SimProp} 
adopting the most recent hadronic model, {Sibyll 2.3d} \citep[][]{Riehn_2020}, an event generator designed for Monte Carlo simulations of atmospheric cascades at ultra-high energies. The hadronic interaction cross section is calculated in the minijet model \cite{PANCHERI1986199}, while the Glauber scattering theory \citep[][]{Glauber:1970jm} is applied in hadron-nucleus collisions and extended with a semi-superposition approach to nucleus-nucleus collisions \citep[][]{Engel:1992vf}. Sibyll 2.3d  allows to compute the cross section for $pp$ and proton-nucleus ($pA$) interactions which, in turn, determines the typical timescale for the spallation process. In addition, Sibyll 2.3d  grants to compute the hadronic interactions taking into account the fragmentation of nuclei and the rapidity of secondary particles produced at each interaction. In particular, the computation of the longitudinal momentum distribution is crucial to determine the fluxes of secondary particles.\\

\begin{figure}
	\includegraphics[width=\columnwidth]{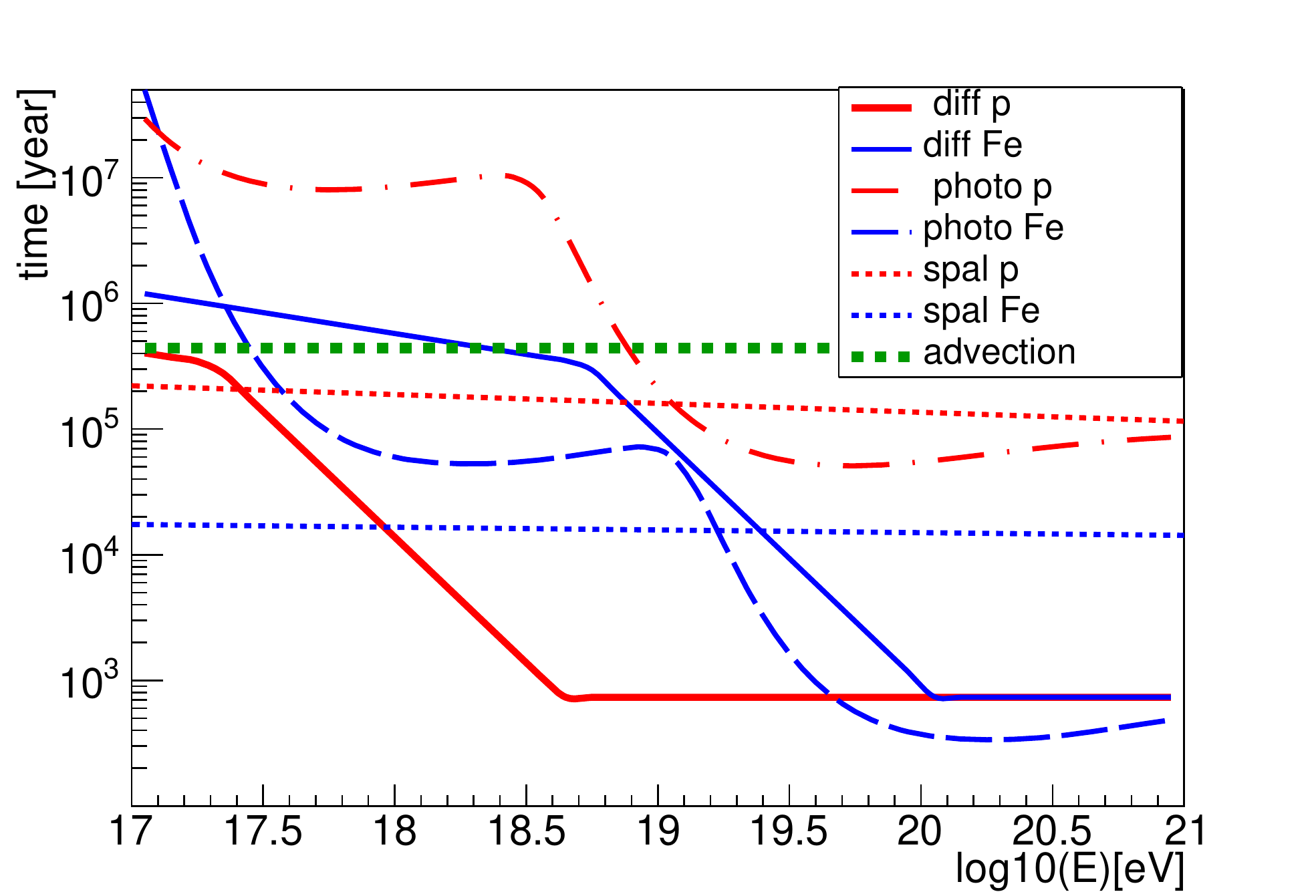}
    \caption{Interaction and escaping timescales for our prototype SBG: photo-hadronic interaction times (dashed-dot lines), spallation times (dashed lines) and diffusion times (solid lines) for protons (red) and Iron nuclei (blue). The green dashed line is the advection time.}
    \label{fig: timescale}
\end{figure}
 On average, high energy particles populating an astrophysical environment are confined for a limited  amount of time before escaping. 
Particles can in fact leave the system as the result of advection in a wind or via diffusion.  
In particular, the advection timescale can be written as $t_{\rm adv}=R/v_{\rm W}$, where $R$ is the source size and $v_{\rm W}$ is the wind speed. The diffusion timescale reads: $t_{\rm D}=R^2/D$, where $D$ is the CR diffusion coefficient computed in the context of quasi-linear theory  and assuming a coherence length $l_c \sim 1 \, \rm pc$ for the magnetic field. 
The expression of the diffusion coefficient is: $D \simeq c r_{L}^{2-\delta} \, l_c^{\delta -1} /3$, where $r_L = E/qB$ is the particle Larmor radius and $\delta$ is the spectral slope of the turbulence, $E$ is the energy and  $q$ is the charge of the particle while $B$ is the strength of the magnetic field.  In particular, we assume $\delta = 5/3$ as prescribed for a Kolmogorov turbulence cascade.
 Following \cite{Subedi:2016xwd}, we additionally consider the transition in the diffusion regime taking place when $r_L \gtrsim l_c$. 
In this energy range the diffusion coefficient is estimated as $D = D_0 (r_L/l_c)^2$, where $D_0$ is the value of the diffusion coefficient computed at the energy $E_0$ such that $r_L(E_0)= l_c$.
We finally estimate the escape timescale, shown in Fig.~\ref{fig: timescale}, as the minimum between the advection and the diffusion time, namely $t_{\rm esc}= {\rm min}[t_{\rm adv},t_{\rm D}]$.\\ 

{Fig.~\ref{fig: timescale} summarizes the typical timescales for interactions and escape in the source environment for our prototype SBG (see Tab. \ref{tab:Reference_SBG}). The interplay between timescales governs the shape of the CR fluxes to be released in the extra-galactic space as well as the mass composition, depending on the source parameters and on the CR spectrum at the acceleration site.}
We observe that, in the lowest energy range ($E \lesssim 10^{18} \, \rm eV$), the spallation has the  shortest timescale, therefore it dominates the transport. 
At higher energies, ($E \gtrsim 10^{18} \, \rm eV$),  the dominant process is the photo-interaction with the infrared photons. Note that in this energy range the diffusion is ballistic and these two process are competing.

Compared to photo-hadronic processes, spallation generates more secondary particles, consequently more photons, neutrinos and a larger number of lighter nuclear fragments.
\begin{figure}
	\includegraphics[width=\columnwidth]{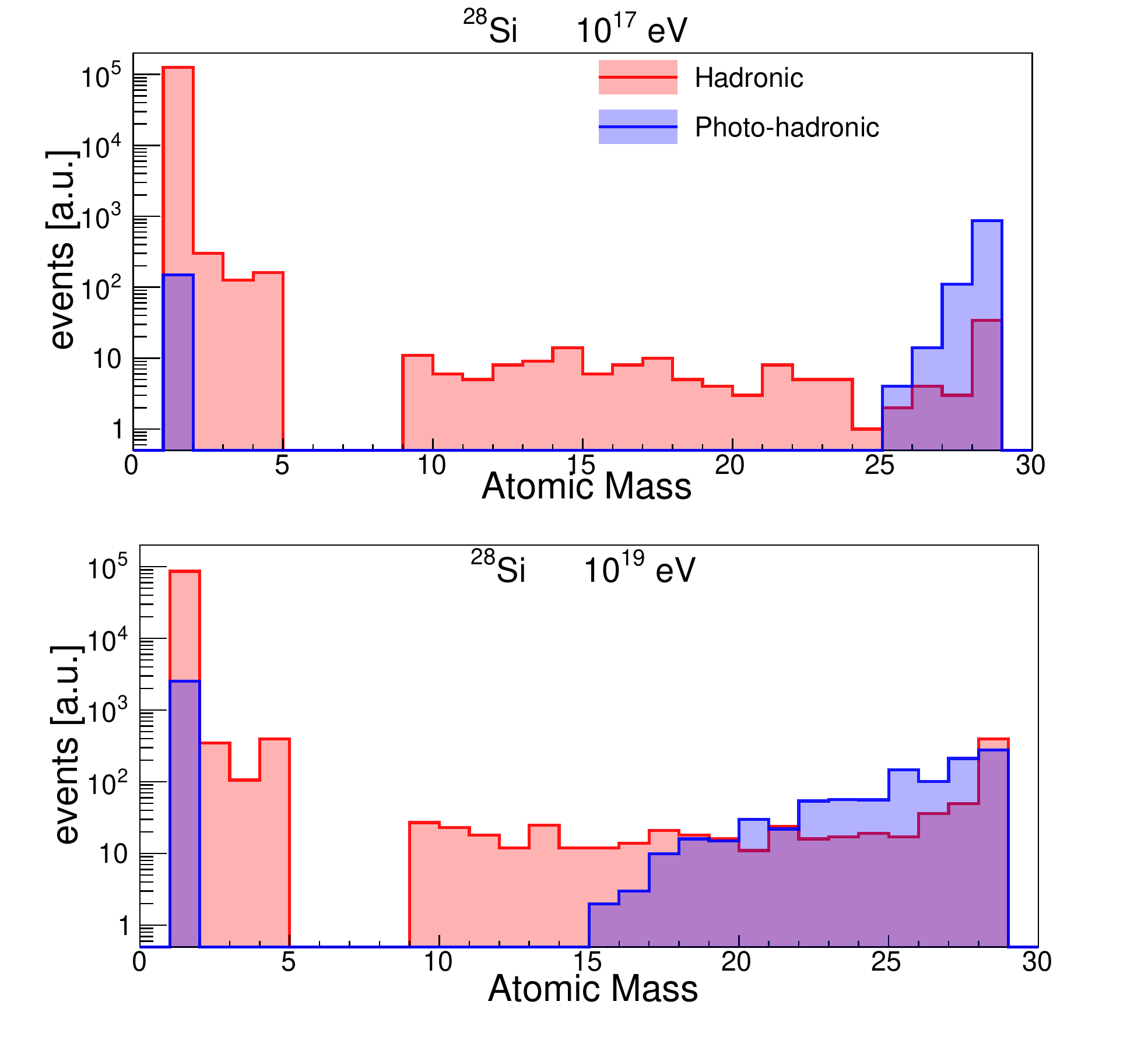}
    \caption{Mass distributions of nuclei escaping from the prototype SBG, for primary Silicon nuclei with energies $10^{17}$ (top) and $10^{19}$ eV (bottom panel). The blue (red) histogram refers to the propagation in the source when only photo-hadronic (hadronic) interactions are permitted.}
    \label{fig:PhVsHadr}
\end{figure}
This different composition of propagated nuclei produces, in turn, a change in the evolution of the nuclear cascade inside the source  environment. Fig.~\ref{fig:PhVsHadr} illustrates the mass distributions of nuclei escaping  our prototype SBG, when the injection is assumed to be characterized only by Silicon nuclei  with energy respectively at $10^{17}$ and $10^{19}$ eV. In particular, we focus on separating the transport effects in presence of a single interaction mechanism inside the source: photo-hadronic-only scenario (blue) is shown separately from the spallation-only one (red). 
 It can be observed that, while the spallation scenario produces efficiently all lighter nuclei, the photo-hadronic scenario does not produce efficiently intermediate mass (C, N, O) nuclei.





\subsection{Implementation of source-interactions in \textit{SimProp}}
\label{Sec: source-int}
In order to estimate the escaping flux from a  SBG, { we develop an extension of a pre-existing} MonteCarlo code {\it SimProp} \citep[][]{Aloisio:2012wj, Aloisio:2015sga, Aloisio:2016tqp, 2017}.
This sofware has been developed and adopted so far in the context of the extra-galactic propagation of UHECRs for instance in \citep[see e.g.][]{combFit, Guido_ICRC2021} while, in this work, it has been modified to model also the transport inside the source. 
The propagation inside the source is performed in the context of a leaky-box model according to the following assumptions: 1) particles are injected in the SBN; 2) all typical timescales are independent on the position; 3) particles escape if the
interaction probability is smaller than the escape one, otherwise they lose energy and all their byproducts are accounted for in the following step of the propagation; 4) particles interacting so often that their energy is not in our range of interest are not propagated anymore.

{\it SimProp} simulates the propagation of UHECRs through the extra-galactic medium assuming a given spectrum 
of injected particles. Note that the propagation in the source depends on the parameters of the source but not on the spectral parameters. For this reason, the in-source propagation is done once for each set of set of source parameters using a unique flux ($ \propto E^{-1}$). When  spectral parameters (spectral index and rigidity cut) are changed the  corresponding ejected spectra can be obtained simply re-weighting the elemental spectra. This procedure
has the advantage of highly reducing the
computational time required to explore the parameter space. For the last step, the propagation from the SBN to the Earth,  we adopt
the same procedure described in appendix A of \cite{combFit}.

\section{Comparison to experimental data}
\label{Sec: 3-Experimental_data}

The aim of this work is to test the hypothesis of SBGs being the sources of UHECRs. Such an investigation is performed by comparing with experimental data the CR flux as modified by interactions in the SBNi and in the extra-galactic propagation. 

{We adopt a measurement of the energy spectrum in $\log_{10}(E/\rm eV)$ bins of 0.1 width from ${17.8}$ eV to ${20.2}$ eV, obtained with the data collected over 15 years with the surface detector array of the Pierre Auger Observatory \cite{PhysRevLett.125.121106}. As for the $X_{\rm max}$ distributions \citep[]{PRDpaper2014}, we consider $\log_{10}(E/\rm eV)$ bins of 0.1 from ${17.8}$ eV to $19.6$ eV{{, plus one additional larger bin containing events with energies above $10^{19.6}$ eV}}; each $X_{\rm max}$ distribution is binned in intervals of 20 $ \rm g \ cm^{-2}$. 
Following \cite{combFit}, we use the deviance $D = - 2\mathrm{ln}(\mathcal{L}/\mathcal{L}_{\text{sat}})$ as estimator of the agreement of our parametric model to data, where $\mathcal{L}$  is our model and $\mathcal{L}_{\text{sat}}$ is a model that perfectly describes the data. The total deviance consists of two terms, $D_{\rm J}$ and $D_{X_\mathrm{max}}$. 
The former refers to the energy spectrum and is a product of Gaussian distributions, the latter is a product of multinomial distributions used for the fit of the $X_{\rm max}$. 
They are modeled as Gumbel distribution functions whose parameters depend on the hadronic interaction model. For the current analysis we adopt EPOS-LHC \cite{Pierog:2013ria} as hadronic interaction model.}\\  


\subsection{Characterization of the parameter space}

%
%
In what follows, we present the  set of free parameters and assumptions adopted for the source-propagation model in the present analysis (see also Tab.~\ref{tab:Param_space_SBG}) characterizing the source environment, the injection parameters of the accelerated CRs and the details of the extra-galactic propagation, as well as the additional parameters needed for the low-energy region of the measured spectrum and composition.

\paragraph*{Source parameters:}

The free parameters associated to the source are the total infrared luminosity of the SBG, $L_{\rm IR}$, and the radius $R$ of the SBN region. 
The former is allowed to range from the typical luminosity of mild nearby SBGs such as M82 or NGC253 ($10^{44} \, \rm erg \, s^{-1}$) up to the value ($10^{46} \, \rm erg \, s^{-1}$) featured by the most powerful ultra-luminous infrared galaxies (ULIRGs) such as Arp 220 \citep[][]{Galliano2007StellarEE}. 
The latter is assumed to vary from a minimum value of 150 pc, as inferred for NGC253 \citep[see e.g.][]{Peretti:2018tmo}, up to 250 pc as a standard value for the scale-height of thin disks in spiral galaxies like the Milky Way \citep[see][]{carroll_ostlie_2017}. 
The luminosity and the size play a key role on the transport of UHECRs. In particular, $L_{\rm IR}$ affects energy losses whereas $R$ {has an impact on both interaction and} escape time.  
The target density $n_{\rm ISM}$ is connected to the star formation rate and in turn to the IR luminosity according to the Kennicutt-Schmidt scaling \citep[][]{Kennicutt:1997Relation}.
Therefore the target density is uniquely determined by the total IR luminosity as: 
\begin{equation}
n_{\rm ISM} \simeq 200 \cdot \bigg[\dfrac{L_{\rm IR}}{L_{\rm IR, M82}} \bigg]^{0.715} \, \rm cm^{-3}
\end{equation}
where the exponent of the correlation is  in agreement with \cite{Kennicutt_2012}.
Finally, the magnetic field $B$ in the SBN is assumed as a fixed parameter at 200 $\mu$G as representative value for SBNi. 
Similarly, the coherence length of the magnetic field fix at 1 pc as plausible value \citep[see e.g.][]{Peretti:2018tmo}.


\paragraph*{Injection parameters:}
We assume that CRs are injected as a power-law spectrum of index $\gamma$, such as the injected flux is proportional to $ E^{-\gamma}$, with maximum rigidity $R_{\rm cut}$. 
In particular, we consider that $\gamma$ ranges from a maximum value $\gamma=2$, as predicted by the diffusive shock acceleration, up to a minimum value $\gamma=1$. 
Hard spectra have been already proposed in the literature \citep[see e.g.][]{Blasi:2000xm,Kotera:2015pya,Unger:2015laa}, in fact they could result from different possible scenarios such as: multiple shocks \citep[see e.g.][]{Pope1994}, relativistic magnetic reconnection \citep[see e.g.][]{Guo_reconnection} or peculiar transport properties encountered by particles before being able to leave the accelerator region.
For the rigidity cutoff we assume the range $10^{18}-10^{19} \, \rm eV$ as suggested by recent results on the combined fit performed by the Pierre Auger Collaboration \citep{combFit}.
For simplicity we work under the assumption of a single heavy nuclear mass $A$ injected in the SBN environment.
Such an assumption allows us to precisely explore the fragmentation of heavy nuclei and the associated production of lighter byproducts. 
In agreement with \cite{Unger:2015laa} and \cite{combFit} we fix $A$ to the atomic mass value of Silicon-28.


\paragraph*{Extra-galactic propagation:}

{\it SimProp} implements two photo-disintegration cross section models: TALYS \citep{TALYS_1, TALYS_2, TALYS_3} and PSB \citep{PSB_1, PSB_2}, and two possible models for the EBL: Gilmore \citep{Gilmore,Gilmore_EBL} and Dominguez \citep{Dominguez}.
In this work we adopted TALYS and Gilmore as photo-disintegration cross section (both for the computation of the interactions in the source environment and in the extra-galactic space) and EBL model respectively.
Finally, we assume that our UHECR sources are distributed in redshift following the star formation rate (SFR) evolution up to $z_{\rm max}=6$. 
The SFR dependence on redshift reads \cite{Y_ksel_2008}:
\begin{equation}
\centering
S_{\rm SFR}(z) \propto \begin{cases}
 (1+z)^{3.4} & z \leq 1 \\
2^{3.7} \cdot (1+z)^{-0.3} &  1  < z \leq 4 \\
2^{3.7} \cdot 5^{3.2} \cdot (1+z)^{-3.5} & z > 4 \\
\end{cases}
    \label{eq:SFR}
\end{equation}    
%

    

%

%


\paragraph*{Low-energy component:}

Similarly to \cite{Unger:2015laa}, we additionally introduce a heavy CR flux below the ankle. This is needed because the disintegration of nuclei in the source environment would produce only light nuclear fragments in the energy range of the ankle, which is not consistent with what is expected from the measured mass composition.
Such a spectral component could be ascribed to a different class of extra-galactic sources \citep[see e.g.][]{Peretti-wind} as well as to rare Galactic PeV-atrons or re-acceleration of Galactic CRs at the Galactic wind termination shock \citep[see e.g.][]{Thoudam_WindShock}. 
In our work we assume this additional component to be with a fixed spectral index $\gamma=4.2$, dominated by the Nitrogen mass group \citep[see also][]{Guido_ICRC2021} while we allow for a free normalization.

%
\begin{table}
    \centering
\begin{tabular}{p{2.7 cm}|p{2.4 cm}|p{1 cm}|p{1.1 cm}} 
\textbf{Parameter} & \textbf{Range} & &\textbf{Best case} \\ \hline
\multicolumn{4}{c}{\textbf{Source parameters}} \\ \hline

{$R/ \rm pc$}  &    [150,250]  & free & 250   \\
{$\log_{10}(L_{\rm IR}/ (\rm erg/s))$}   &    [44,46] & free &  44.7  \\
{$B/ \rm \mu G$}  &    -  & fixed & 200   \\
{$l_{\rm c}/ \rm pc$}  &    -  & fixed & 1   \\
\hline
\multicolumn{4}{c}{\textbf{Injection parameters}} \\ \hline
{$\gamma$}   & [1,2] & free & 1  \\
{$\mathrm{log_{10}}(R_{\rm cut}/ \rm V)$} &   [18,19]  & free & 18.5 \\
{$A$}   &  & fixed & 28  \\
\hline
\multicolumn{4}{c}{\textbf{Extra-galactic propagation's parameters}} \\ \hline
{Ph-dis. cross sect.}  & -  & fixed & TALYS  \\
{EBL model}   & -  & fixed & Gilmore  \\
{Evolution} & -  & fixed & SFR \\
\hline
\multicolumn{4}{c}{\textbf{ Low-Energy component parameters}} \\ \hline
{Spectral index}  &  -  & fixed & 4.2  \\
{Mass}   &  -  & fixed & 14  \\
\hline
\end{tabular}
 \caption[Parameter space for prototype case]{Parameter space for the prototype model.}
    \label{tab:Param_space_SBG}
\end{table}

\section{Results}
\label{Sec: 5-Results}

Starting from the set of parameters of our prototype SBG (see Tab.~\ref{tab:Reference_SBG}), which is representative of the most common class of mild starbursts (including nearby sources such as M82 and NGC253), we perform a parameter space scan in order to find the best configuration. The choice of mild starbursts as starting point of our investigation was suggested by the form of the star-formation-rate function, which suggests that M82-like galaxies are the most common and abundant in the local Universe.
The best source parameters are shown in the last column of Tab.~\ref{tab:Param_space_SBG} and correspond to a luminosity $\sim 5$ times higher than our starting prototype; such a luminosity is typical of a more active class of galaxies known as luminous infrared galaxies (LIRGs). 
SBGs with these properties, given the shape of the luminosity functions \cite{2013Gruppioni}, are somehow less common compared to the prototype model. On the other hand, LIRGs do not occupy the highest end of the luminosity function where ultra-LIRGs (ULIRGs) such as Arp220 can be found. Therefore, one could speculate that possibly galaxies with an infrared luminosity above a certain threshold are likely to host UHECR acceleration sites.

\begin{figure}
	\includegraphics[width=\columnwidth]{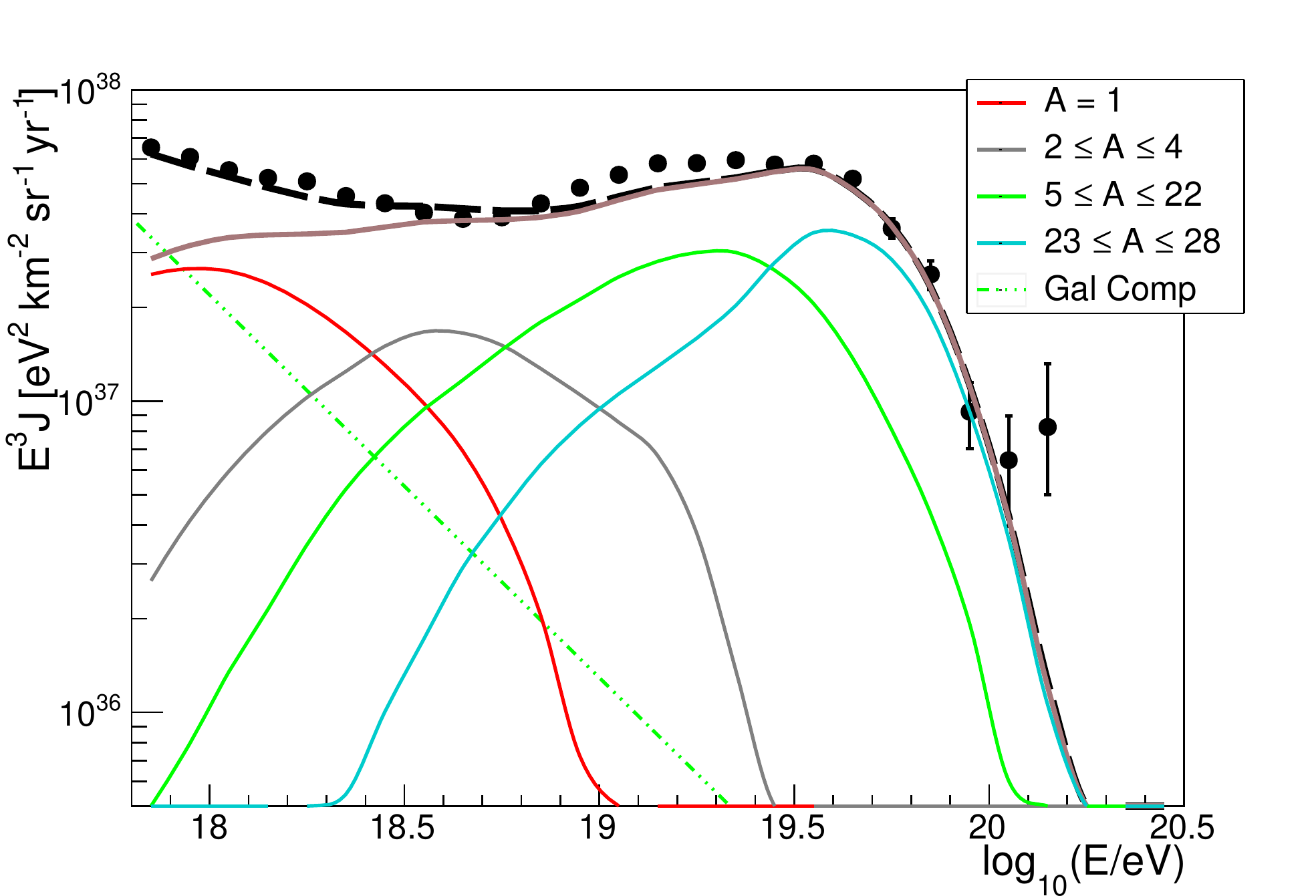}
		\includegraphics[width=\columnwidth]{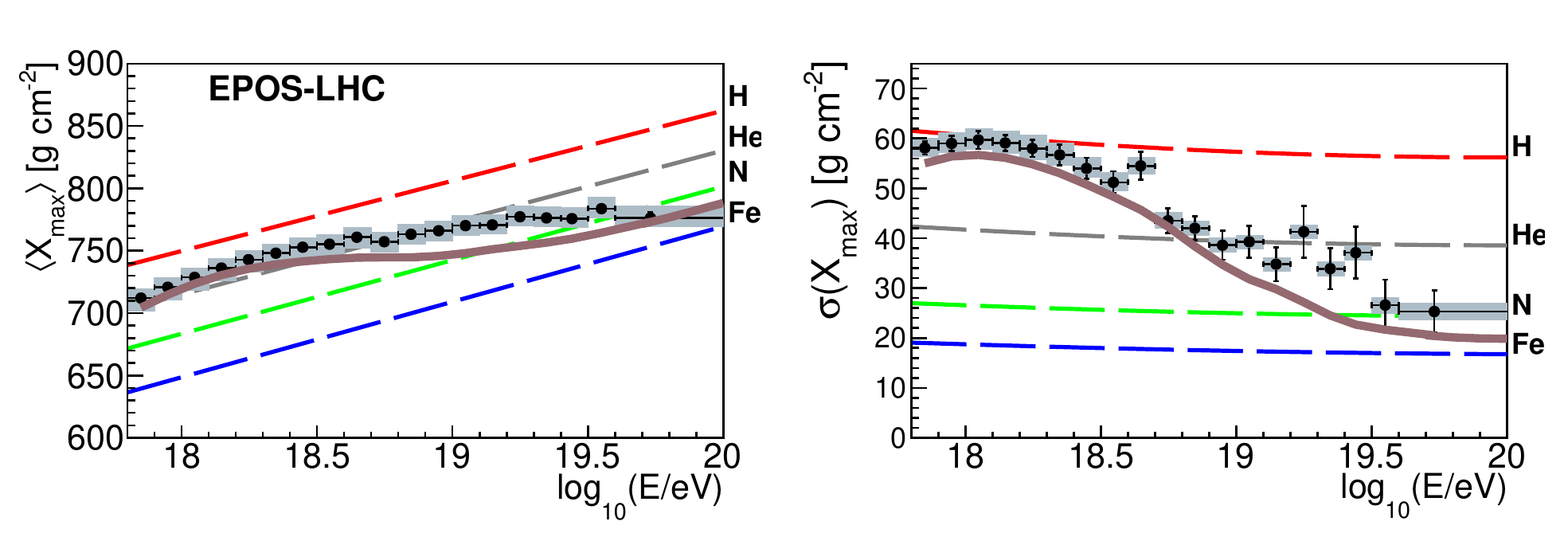}

    \caption{Top: All-particle best-fit scenario and partial spectra related to different detected mass groups compared to the measured spectrum \cite{Spectrum_2019}.
    Bottom: Average (left panel) and standard deviation (right panel) of the experimental (black dots as reported in \cite{Composition_2019}) and expected (lines) $X_{\mathrm{max}}$ distribution.}
    \label{fig:Spectrum_best}
\end{figure}

Fig.~\ref{fig:Spectrum_best} illustrates the spectrum and mass composition of UHECRs at Earth relative to the best fit parameters.
%
%
{Despite the simplicity of our model, our best fit can qualitatively reproduce the ankle feature with a precision of the order of $\sim 10 \%$ (at the energy of the ankle); note also that} our calculation results in a complex evolution of the mass composition with energy that well approximates the ankle feature, while the hierarchical order of the partial fluxes follows the first two moments of the $X_{\rm max}$ distributions. 
In particular, at low energies the composition is dominated by light secondary nucleons whereas, at high energies, the composition becomes heavier due to the dominant escape compared to the interaction rate. 
The additional low-energy component is dominant only below $\sim 10^{18} \, \rm eV$ whereas, at higher energies, the contribution from SBGs dominates. We stress that our results have been obtained under the assumption of a single injected  nuclear species  and a unique stereotypical SBG representative for the whole class instead of a more appropriate luminosity function. The impact of such assumptions is discussed in section \ref{Sec: 6-Discussion-Conclusions}.

The number of SBGs required to describe the data should not exceed the number of such galaxies observed in the local Universe. Comparing the model prediction with data, it is possible to infer the required emissivity $\varepsilon$ needed to power the UHECRs at redshift $z=0$ as:
\begin{equation}
    \varepsilon = \int_{E_{\rm min}}^{\infty} J_{\rm inj} (E) \ E \  dE  \,
\end{equation}
where $J_{\rm inj}$ is the spectrum injected at the source, before considering the interactions, and $E_{\rm min}$ is an arbitrarily low energy value (here $E_{\rm min} = 10^{17}$ eV).
As already discussed in \cite{Condorelli:2021zst}, we define $\alpha$ as the ratio between the CR luminosity (${L}_{\rm CR}$) and IR luminosity  obtained from the best fit ($L_{\rm IR} = 1.2 \cdot 10^{45} \ \rm erg/s $). 
The number density of sources, $n_{\rm SBG}$,  can be estimated as
\begin{equation}
     n_{\rm SBG} = \dfrac{\varepsilon_{\rm CR}}{\alpha \cdot {L}_{\rm IR}} =  5.1 \cdot 10^{-5} \, \Big[\frac{\alpha}{0.1} \Big]^{-1} \ \rm Mpc^{-3} ,
          \label{eq:emissivity_result}
\end{equation}
where $\alpha$ is normalized to $0.1$ under the assumption of a sub-equipartition of the UHECR population compared to the background photon fields.

In \cite{2013Gruppioni}  the luminosity density is computed as a function of the redshift using the luminosity functions.
By integrating these functions for luminosity above the best fit one , we find:
\begin{equation}
     n_{\rm SBG} \simeq 3.3 \cdot 10^{-4} \ \rm Mpc^{-3}
     \label{eq:gruppioni}
\end{equation}
It can be observed that the number density of sources inferred from the integral of the luminosity function (eq. \ref{eq:gruppioni}) is an order of magnitude higher than the one obtained with our model (eq. \ref{eq:emissivity_result}), assuming a fiducial value $\alpha=10 \%$. This result is encouraging since it can be reconciled with a low  dynamical impact of UHECRs in the starburst environment.

\subsection{Exploring the parameter space}

Starting from our best fit scenario (see Tab.~\ref{tab:Param_space_SBG})  we explore the implications  of different parametric configurations in the injection and in the main properties of the stereotypical SBG.

\begin{figure}
	\includegraphics[width=\columnwidth]{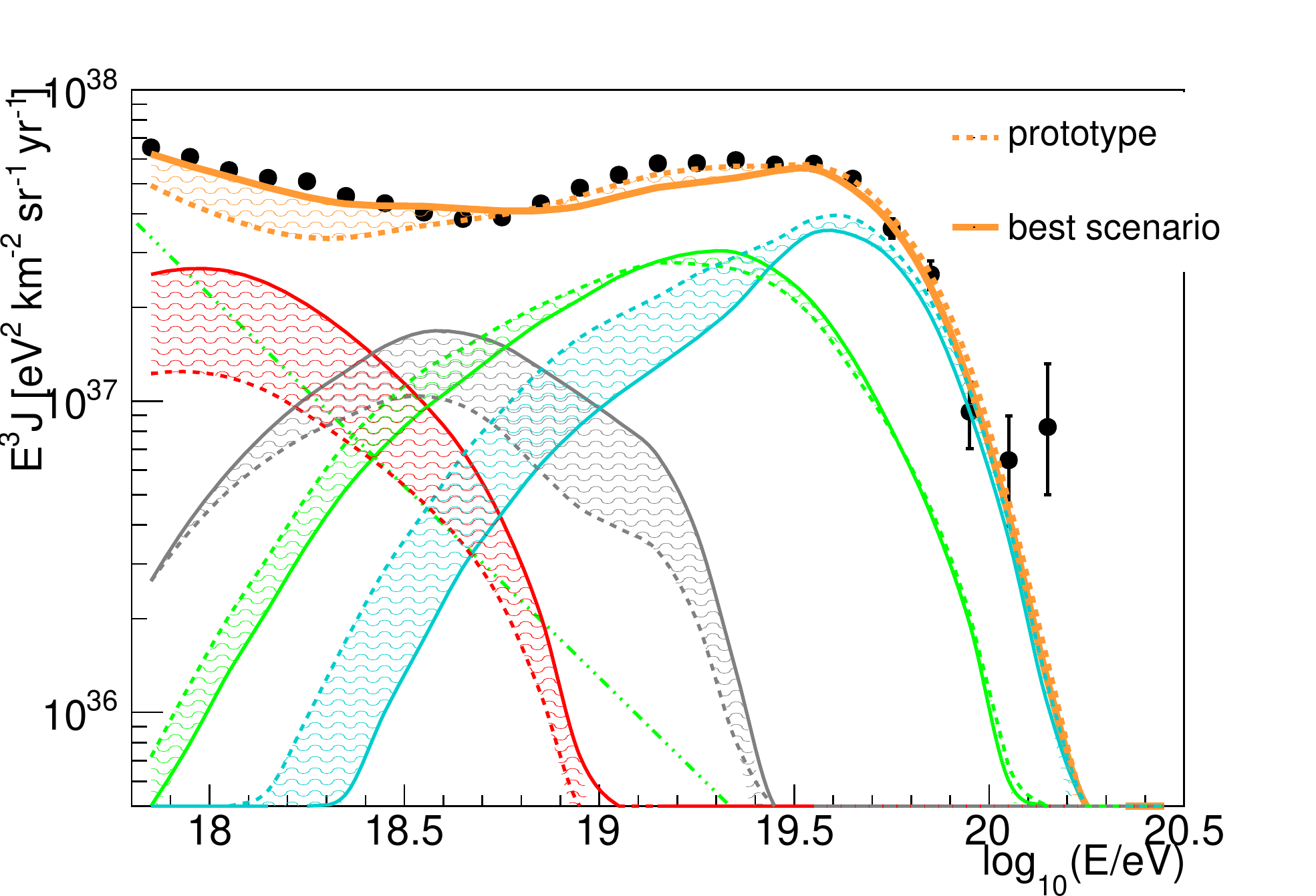}
    \caption{All-particle best-fit scenario and partial spectra related to different detected mass groups corresponding to two different densities of target in the source environment: our prototype (dashed line) and our best-fit scenario, which corresponds approximately to five times the luminosity of M82 (solid line). The shaded area is drawn in order to highlight the differences.}  
    \label{fig: spectrum_source}
\end{figure}

The effects on the spectra at Earth of a different assumption in the IR luminosity, and consequently the ISM density (following the Kennicutt relation \cite{Kennicutt:1997Relation}), are shown in Fig.~\ref{fig: spectrum_source}.
In particular, one can appreciate the difference in the total flux (thick and dashed orange lines) and the associated mass group components, when our best fit (see Tab.~\ref{tab:Param_space_SBG}) or a prototype SBG, as described in Tab.~\ref{tab:Reference_SBG}, is assumed. It is straightforward to notice that the higher the ISM and photon density, the higher is the rate of interactions inside the source. This leads to a higher efficiency of disintegration of Silicon nuclei, thereby producing a larger number of light secondaries and reducing the flux of heavy nuclei.
On the other hand, in the case of ultra luminous infrared galaxies (ULIRGs), characterized by an activity $\simeq 50$ times greater than the prototype SBG, the density target is so high that most of the particles above the ankle are disintegrated, so that the escaping flux cannot reproduce the features of the energy spectrum and the mass composition at the highest energies.

\begin{figure}
	\includegraphics[width=\columnwidth]{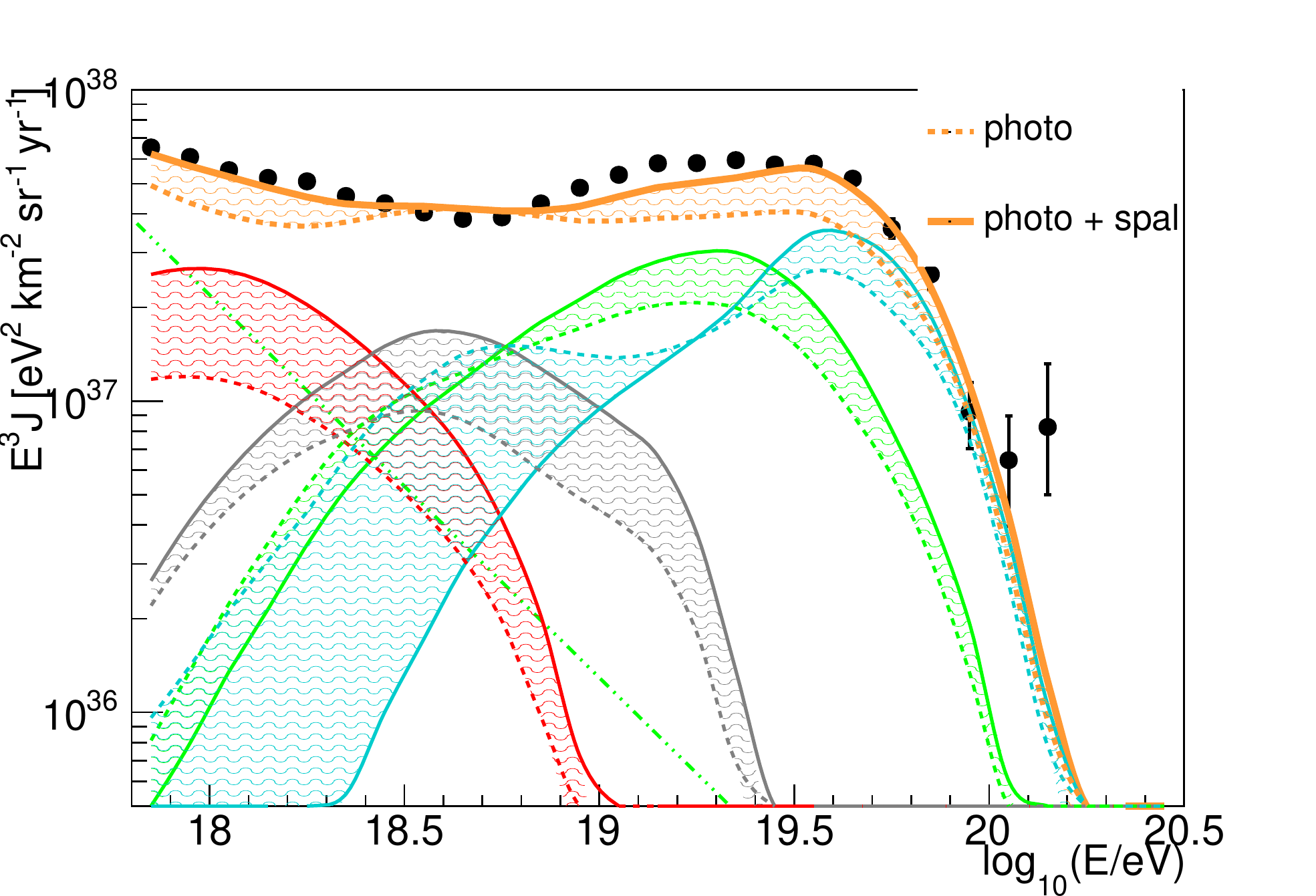}
    \caption{All-particle best-fit scenario and partial spectra related to different detected mass groups corresponding to two different hypotheses: neglecting (dashed line) or including (solid line) hadronic processes in the source environment.}
    \label{fig: spectrum_photo_spal}
\end{figure}
\begin{figure}
	\includegraphics[width=\columnwidth]{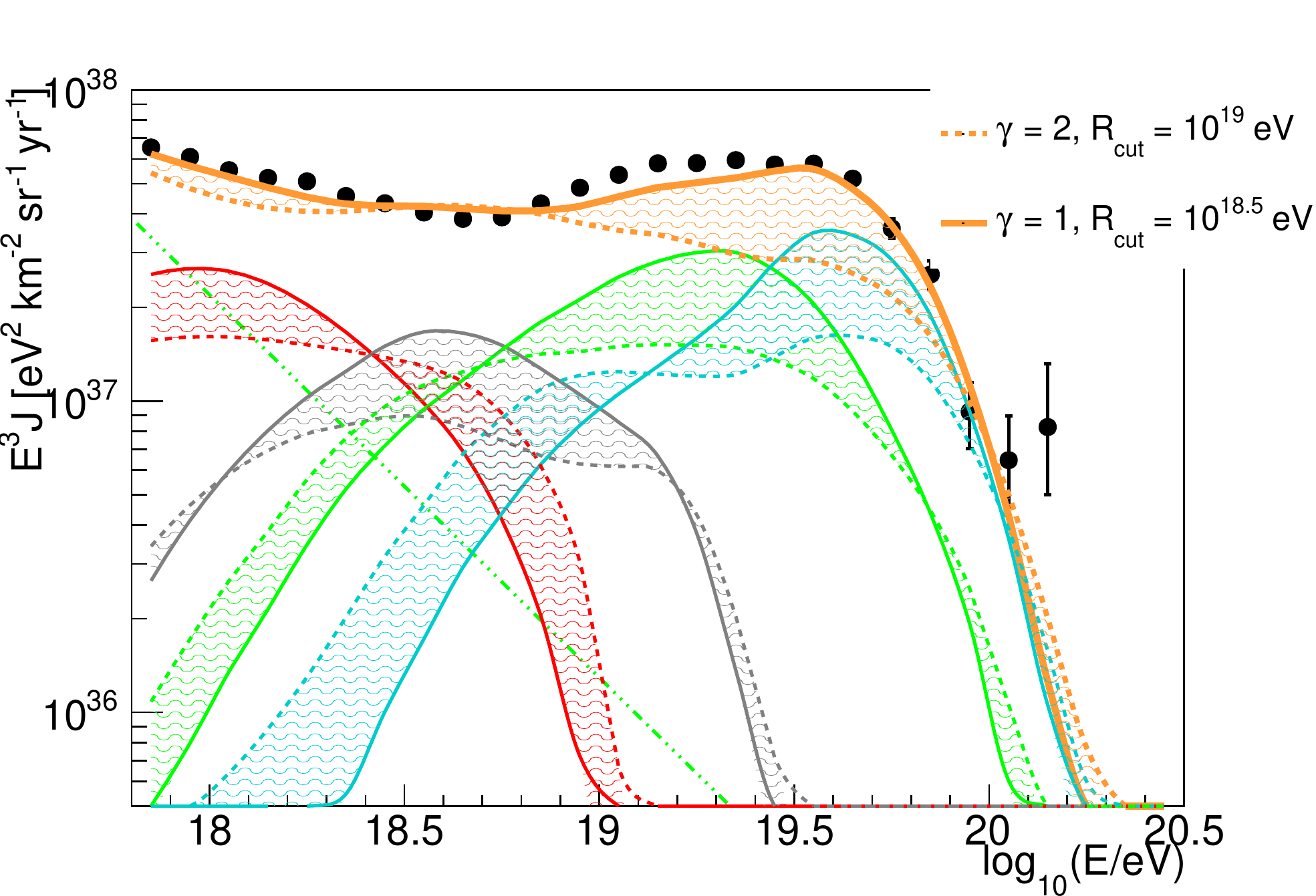}
    \caption{All-particle best-fit scenario and partial spectra related to different detected mass groups corresponding to two different acceleration hypothesis: hard injection spectrum ($\gamma = 1$, solid line) and soft injection spectrum ($\gamma = 2$, dotted line).}
    \label{fig: spectrum_acc}
\end{figure}
In Fig.~\ref{fig: spectrum_photo_spal} the impact of spallation processes in the SBG environment on the  spectrum at Earth is shown, by computing the expected fluxes at Earth with the same parameters as the best fit case and neglecting the spallation in the source environment. The net effect of considering spallation is to increase the efficiency of the disintegration of nuclei with the consequent production of light fragments. However, due to the energy region where the spallation effects are dominant, the effect of neglecting it are stronlgy visible at intermediate to low energies.

UHECRs are assumed to be injected in the SBN environment according to a power-law spectrum of index $\gamma$ and maximum rigidity $R_{\rm cut}$.
In Fig.~\ref{fig: spectrum_acc}, the outcomes of our best fit model ($\gamma=1$, $R_{\rm cut}=10^{18.5}$ eV), are compared with the results obtained under the assumption of a softer injection, ($\gamma=2$, $R_{\rm cut}=10^{19.0}$ eV), as prescribed by the standard diffusive shock acceleration.
We observe that the assumption $\gamma=2$ as well as spectra softer than $\gamma =1$ are disfavoured by our analysis.
%
{As one can see from the} dashed lines, a qualitative description of the spectrum fails especially at the highest energies, where the transport is regulated by the competition between photo-hadronic interactions and diffusion. 

 

\subsection{Constraining the neutrino flux}



Together with the escaping flux of UHECRs, we keep track of the hadronic and photo-hadronic byproducts such as gamma rays and neutrinos. 

The SBN environment is highly opaque to gamma rays with energy $\gtrsim 10 \, \rm TeV$ and, due to the strong magnetic fields typical of such environment, HE electron-positron pairs are expected to cool via synchrotron instead of initiating cascades leading to possible spectral features in the TeV range \citep[][]{Peretti:2018tmo}. 
Therefore, one does not expect a relevant gamma-ray counterpart associated to the presence of UHECRs in the SBN environment. 
For this reason, we leave the investigation on the multi-wavelength consequences of UHECRs in SBN environment to upcoming works.
Different from gamma rays, neutrinos travel practically undisturbed once they are produced.
We compute the production of HE neutrinos both inside the SBG environment and in the propagation of UHECRs from their sources to the Earth. 
Given their low interaction cross section, we only account for the adiabatic energy loss effect on the neutrino flux due to the expansion of the Universe. 
We finally assume an average flavor composition at Earth (1:1:1) after the oscillation through cosmological distances.

%
\begin{figure}
	\includegraphics[width=\columnwidth]{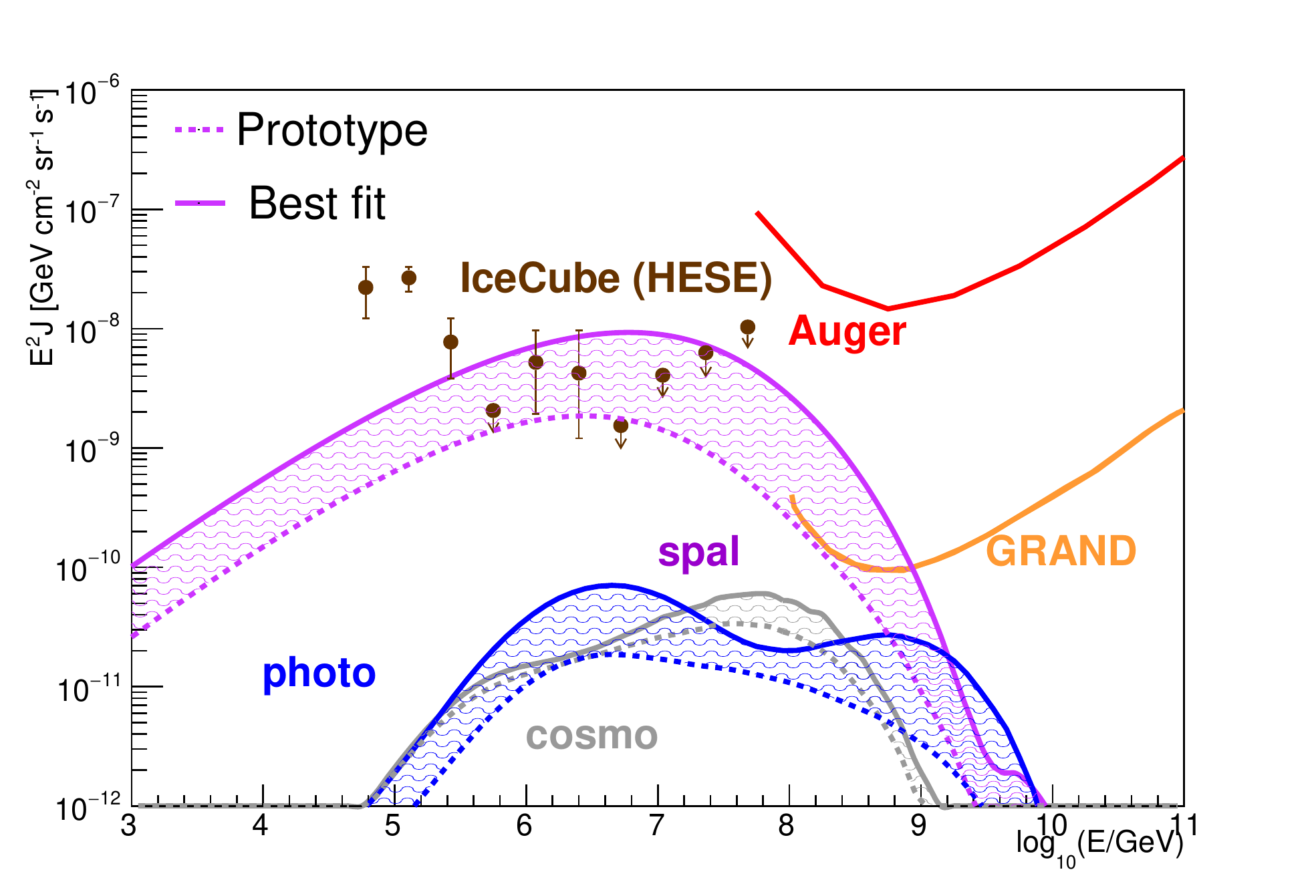}
    \caption{Single-flavour neutrino fluxes for the prototype case (dashed) and best case (solid), compared to the IceCube neutrino flux \cite{abbasi2020icecube}, the expected sensitivity of GRAND after three years of operation \cite{Alvarez-Muniz:2018bhp} and to the limits for the cosmogenic neutrinos by the Pierre Auger Collaboration \cite{Aab_2019}. The shaded area is drawn in order to highlight the difference}
    \label{fig: neutrinos with lines}
\end{figure}

In Fig.~\ref{fig: neutrinos with lines} we show the diffuse neutrino fluxes associated to the different contributions considered in this work: cosmogenic neutrinos (grey lines), namely those neutrinos produced by the interaction of UHECRs with the Cosmic Microwave Background (CMB) and the Extragalactic Background Light (EBL), the neutrinos produced by photo-hadronic interactions of UHECRs in the source (blue lines) as well as the neutrinos produced by hadronic interactions of UHECRs in the source (magenta lines). 
The neutrino flux resulting from our calculation is compared with the Auger limit, the limit expected by GRAND \citep[][]{Alvarez-Muniz:2018bhp} after three years of operation and the flux observed by IceCube \citep{abbasi2020icecube}.
Two different model predictions, solid and dashed lines, are shown and compared in Fig.~\ref{fig: neutrinos with lines}. They refer respectively to our  best fit SBG and the prototype SBG (M82-like assumption). The corresponding UHECRs are reported in Fig.~\ref{fig: spectrum_source}. 
Comparing the two model predictions, it is possible to observe how the source-neutrino fluxes increase with increasing infrared luminosity (as for instance already reported in \cite[]{Biehl:2017zlw,Boncioli:2018lrv}) and gas density, as shown already in \cite{Muzio:2021zud}. 
On the other hand, cosmogenic neutrinos are almost unaffected by the source properties we focused on, being mostly related to the spectral characteristics of UHECRs escaping their sources, such as the spectral index and the maximum rigidity. 
We finally notice that the expected cosmogenic neutrinos are way below the current limits.

It is interesting to notice that the source-neutrinos produced in $p\gamma$ interactions are almost comparable to the expected cosmogenic fluxes in the presented cases, while the source-neutrinos produced in $pp$ and $pA$ interactions dominate the neutrino flux detected by IceCube. 
This result suggests that the flux of neutrinos observed by IceCube above $\sim 100 \, \rm TeV$ could be a direct consequence of the presence of UHECR accelerators in SBGs. Such a multi-messenger connection implies that we can possibly investigate the sources of UHECRs by looking at the galaxies shining in neutrinos of energy at around 1-100 PeV.

\section{Discussion and conclusions}
\label{Sec: 6-Discussion-Conclusions}

In this work  we develop a source-propagation model in order to explore whether SBGs can be the sources of UHECRs studying in detail the interactions taking place in the SBG environment and in the propagation to the Earth. In particular, we compute for the first time proton-proton and proton-nucleus interactions in the SBG environment and we analyse the impact to the UHECR and neutrino fluxes in addition to photo-hadronic interactions.

We work under the assumption that the sources of UHECRs are all characterized by some representative properties, and we compare our model prediction with the energy spectrum and mass composition measured by the Pierre Auger Observatory. 
We show that SBGs can qualitatively well describe the measured UHECR spectrum and composition.
We also compute the neutrino flux associated to the transport of UHECRs and we show how this improves the constraining capability of our model; this could allow us to consider whether a set of parameters at the source can describe the UHECR data without overshooting the measured neutrino fluxes. 
In particular, we find that, if SBGs were hosting the UHECRs accelerators, they would provide a sizeable contribution to the neutrino flux observed by IceCube at energies $\gtrsim 10^2 \, \rm TeV$. We find that the expected neutrino flux from sources is
strongly predominant with respect to the cosmogenic one. In addition,
we show that hadronic interactions could be crucial for explaining the
measured neutrino flux, and deserve more detailed studies applied to
source environments, since they could be able to hide the outcome in neutrinos from photo-hadronic interactions in the source environnment and in the extra-galactic space.
Therefore, the detection of steady PeV neutrinos emitters could guide in the near future in a complementary search of the sources of UHECRs. 

Within the explored parameter space, we find that the data can be well described if the UHECR nuclei are injected with a hard spectrum. We stress here that, even though the standard injection $\sim E^{-2}$, typical of the diffusive shock acceleration, is disfavoured, this does not necessarily rule out such a process from accelerating the UHECRs. Hard spectra could be in fact obtained in the context of diffusive shock acceleration through several possible conditions that could be realized in the core of SBGs such as: multiple shocks, converging flows, particle reacceleration or transport conditions in the acceleration region which differ from the ones in the whole SBG.

The results here presented have been obtained under the assumption of a single injected heavy nuclear species. An investigation of scenarios where multiples masses are injected with different relative composition, which might also soften the spectral index found at the injection, goes beyond the main goal of this work and is left for future investigations. Nevertheless, we highlight that the assumed scenario fails when the injected masses are heavier than silicon nuclei. 
In particular, we tested the  scenario of iron nuclei at injection, and we found that the description of data considerably worsens. 
The reason for this is twofold: 1) the expected composition at Earth is too heavy compared to the  observational results from the Pierre Auger Observatory; 2) the maximum rigidity of the Iron nuclei is defined by the comparison of the expected UHECR spectrum to the measured one, and the maximum energy of the nucleons from the disintegration (being $1/A$ of the maximum energy of nuclei) is too small to well describe the data at the energy of the ankle.

Several hypotheses have been explored regarding the parameters of the sources, such as different luminosities of the prototype SBG and different spectra of injected particles. 
Considering a source with standard properties is clearly a limitation for this analysis; on the other hand, this approach can highlight the possibility of the existence of some average properties characterizing a class of UHECR sources.
An interesting  improvement of the current work could rely on the use of a catalogue or the luminosity functions of galaxies instead of a single prototype. 
In fact, it is natural to expect that SBGs with different luminosities could contribute at different level to the energy spectrum and probably better describe the Auger data. 

%

%
The HE neutrino flux could serve as multimessenger constraint for the sources of UHECRs. A neutrino flux exceeding the measured one improves the constraining capability of UHECR data; {this represents a powerful tool if compared to other models in literature \cite{Anchordoqui_SBG}, where the neutrino flux accompanying the cosmic rays is extremely suppressed}. Nonetheless, it is also important to notice that
in this work the assumption of all identical sources distributed in the
Universe has been considered up to a redshift $z=6$. This hypothesis affects more the expected neutrino fluxes rather than the UHECR fluxes, which are expected to be originated not far than $z = 1$. 
In addition, the use of  luminosity functions instead of a single prototype is expected to lower and widen the  neutrino fluxes.  

Future improvements of this work shall also include the production and propagation of photons inside SBGs since additional multiwavelength constraints could be found, especially in the hard X-rays, where electrons and positron pairs are likely to emit via synchrotron on the strong magnetic fields typical of SBGs. 
As the neutrino fluxes, the expected photon fluxes can be compared to experimental data, thereby improving the constraining capability of our model. 

Finally, further improvements of source-propagation analyses can be expected by the increasing precision in the determination of the mass composition at the highest energies as expected by the upgrade of the Pierre Auger Observatory \citep{PierreAuger:2021ccl}.


\section*{Acknowledgements}
AC, DB and SP acknowledge their participation to the Pierre Auger Collaboration. The authors would like to thank Felix Riehn, for help in including Sibyll2.3d in {\it SimProp}, and Francesco Salamida for his support in the use of {\it SimProp} in different stages of the analysis. AC gratefully acknowledges funding from ANR via the grant MultI-messenger probe of Cosmic Ray Origins (MICRO), ANR-20-CE92-0052.
The research activity of EP was supported by Villum Fonden (project n. 18994) and by the European Union’s Horizon 2020 research and innovation program under the Marie Sklodowska-Curie grant agreement No. 847523 ‘INTERACTIONS’.


\bibliography{main}

\end{document}